\documentclass[aps,prl,groupedaddress,twocolumn,superscriptaddress]{revtex4-1}
\usepackage{graphicx}
\usepackage{nicefrac}
\usepackage{textgreek}
\begin{document}

\title{High-precision measurement of the proton's atomic mass}
\author{F. Hei\ss e}
\affiliation{Max-Planck-Institut f\"ur Kernphysik, Saupfercheckweg 1,
69117 Heidelberg, Germany}
\affiliation{GSI Helmholtzzentrum f\"ur Schwerionenforschung GmbH,
Planckstra\ss e 1, 64291 Darmstadt, Germany}
\author{F. K\"ohler-Langes}
\affiliation{Max-Planck-Institut f\"ur Kernphysik, Saupfercheckweg 1,
69117 Heidelberg, Germany}
\author{S. Rau}
\affiliation{Max-Planck-Institut f\"ur Kernphysik, Saupfercheckweg 1,
69117 Heidelberg, Germany}
\author{J. Hou}
\affiliation{Max-Planck-Institut f\"ur Kernphysik, Saupfercheckweg 1,
69117 Heidelberg, Germany}
\author{S. Junck}
\affiliation{Institut f\"ur Physik, Johannes Gutenberg-Universit\"at,
55099 Mainz, Germany}
\author{A. Kracke}
\affiliation{Max-Planck-Institut f\"ur Kernphysik, Saupfercheckweg 1,
69117 Heidelberg, Germany}
\author{A. Mooser}
\affiliation{RIKEN, Ulmer Fundamental Symmetries Laboratory, Wako,
Saitama 351-0198, Japan}
\author{W. Quint}
\affiliation{GSI Helmholtzzentrum f\"ur Schwerionenforschung GmbH,
Planckstra\ss e 1, 64291 Darmstadt, Germany}
\author{S. Ulmer}
\affiliation{RIKEN, Ulmer Fundamental Symmetries Laboratory, Wako,
Saitama 351-0198, Japan}
\author{G. Werth}
\affiliation{Institut f\"ur Physik, Johannes Gutenberg-Universit\"at,
55099 Mainz, Germany}
\author{K. Blaum}
\affiliation{Max-Planck-Institut f\"ur Kernphysik, Saupfercheckweg 1,
69117 Heidelberg, Germany}
\author{S. Sturm}
\affiliation{Max-Planck-Institut f\"ur Kernphysik, Saupfercheckweg 1,
69117 Heidelberg, Germany}

\date{\today}

\begin{abstract}
We report on the precise measurement of the atomic mass of a single proton
with a purpose-built Penning-trap system.
With a precision of 32 parts-per-trillion our result not only improves on the current
CODATA literature value by a factor of three, but also disagrees with it at a level of 
about 3 standard deviations.
\end{abstract}


\maketitle

The properties of the basic building blocks of matter shape a network of
fundamental parameters, which are crucial to develop precise quantitative
understanding of nature and its symmetries. One of these fundamental
constants is the mass of the proton $m_{p}$, which has always been a target
and yardstick of precision experiments
\cite{PhysRevLett.73.1481,Borgenstrand436573,
doi:10.1063/1.57450,1402-4896-66-3-002,Solders232785}.
It is thus correlated with most other parameters of atomic physics. For
example, its value influences the Rydberg
constant~\cite{RevModPhys.88.035009}, and it is also required for the precise
comparison of the masses of the proton and antiproton, in order to perform a
stringent test of CPT invariance via a hydrogen anion \cite{Ulmer2015}.

All recent proton mass values are based on Penning-trap measurements, where the
cyclotron frequencies $\nu_c=\frac{1}{2\pi}\frac{q}{m}B$ of the proton (or
$\text{H}_2^+$) and a reference ion with respective charge-to-mass ratios
$\nicefrac{q}{m}$ are compared in the same magnetic field $B$.
In this letter we report on a high-precision measurement of $m_{p}$ in atomic
mass units, which is based on cyclotron frequency comparisons of protons and
highly charged carbon ($^{12}\text{C}^{6+}$) ions.
While the largely different charge-to-mass ratio between the proton and the
$^{12}\text{C}^{6+}$ ion imposes technical challenges to be discussed later,
the comparison with the atomic mass standard allows us to determine the mass
of the proton directly in atomic mass units. In order to do so, we have to
relate the mass of the $^{12}$C$^{6+}$ ion to that of a $^{12}$C atom: 
\begin{equation}
m(^{12}\text{C}^{6+})=
m(^{12}\text{C})-6 m(e^-)+\sum_{i=1}^{6}\frac{E_{b,i}}{c^2}. \label{iE}
\end{equation}   
Here, $E_{b,i}$ denotes the binding energies of the six removed electrons,
$c$ the speed of light in vacuum and $m_e$ the electron
mass \citep{Nature2014}. Since $m(^{12}\text{C})$ is $12~\text{u}$
by definition and the atomic mass of the electron has been previously
determined by our group with $2.9\times 10^{-11}$ relative uncertainty, this
relation is limited only by the knowledge of electronic binding energies.
The currently tabulated values in the NIST table of ionization energies
\cite{NIST_IP} allow to derive
$m(^{12}\text{C}^{6+})=11.996\,709\,626\,413\,9(10)~\text{u}$
with a relative precision of
$0.08~\text{parts-per-trillion (ppt)}$, which does not pose any
limitation on the precision of the proton's atomic mass reported here.

The measurements have been carried out in a highly-optimized, purpose-built
cryogenic Penning-trap setup, dedicated to mass measurements on light ions,
which is a successor experiment of the Mainz $g$-factor experiment for highly
charged ions \cite{PhysRevLett.107.023002,PhysRevLett.110.033003,Nature2016}.
While the superconducting magnet and the experiment's liquid helium cryostat
have been re-used, both the trap section as well as the cryogenic electronics
and detection circuitry have been newly developed. This was necessary to adress
the specifically strong requirements on the quality of the trapping fields,
set by the low mass and charge of the proton and
resulting large motional amplitudes.
\begin{figure*}
	\includegraphics[width=0.75\textwidth]{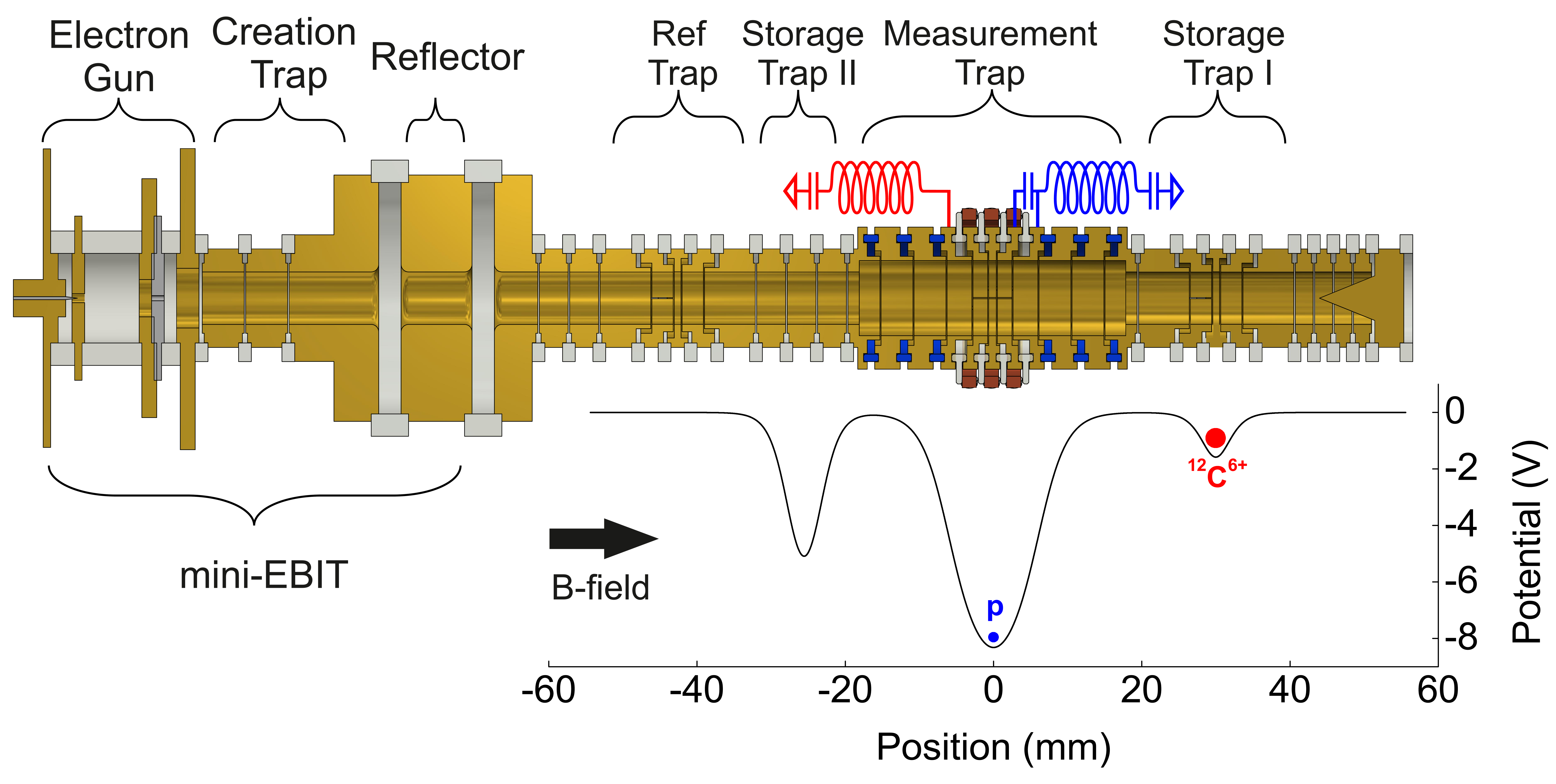}
	\caption{\label{TrapSetup} Sketch of the trap setup.
	The trap tower includes two separate storage traps (ST-I, ST-II),
	the measurement trap (MT) and a reference trap (RT) for magnetic field
	monitoring, which is presently not used. Ions are created in-situ using
	a mini-EBIT \cite{1742-6596-58-1-021}. By shuttling the ions between the
	storage traps and the MT, the time between successive measurements is
	minimized. Individual superconducting detection circuits for the proton
	(blue) and for the carbon ion (red), allow measurements at the identical
	electrostatic field configurations and thus guarantee the identical position and magnetic
	field. For details see text.}
\end{figure*}

The highly charged $^{12}\text{C}^{6+}$ ion as well as the proton are created
in the hermetically sealed and cryogenically cooled trap chamber using an
integrated miniature electron beam ion source (EBIS)
\cite{1742-6596-58-1-021,0953-4075-43-7-074016}, which ablates atoms from a
carbon nanotube-filled PEEK target (TECAPEEK \cite{Tecapeek}). After creation,
the ions are shuttled to the measurement trap (MT, see Fig.~\ref{TrapSetup})
by adiabatically shifting the electrostatic trapping potential along the
magnetic field axis. Here, all ions except for one proton or one
$^{12}\text{C}^{6+}$ ion, respectively, are ejected from the trap. After that,
the ion of interest is transported to a
\textquotedblleft storage\textquotedblright{}
trap (ST-I, see Fig.~\ref{TrapSetup}). Next, a new cloud
of ions is produced in the EBIS and the process is repeated to place the second
ion in the MT. Storing both ions of interest simultaneously in the same trap
setup but at different locations allows rapid swapping of the ions between the
MT and one of the two STs, and thus drastically reduces the time between
measurements of the two ions \cite{Ulmer2015}. Additionally, the creation process 
can change the effective electric potential by charging unavoidable non-conductive 
patches on the trap electrodes also referred to as patch potentials, 
which is also avoided by our method since no reloading is required. Cryopumping in the 
sealed chamber provides a virtually perfect vacuum
of better than $10^{-17}~\text{mbar}$, which prevents any unwanted interaction
of the ion of interest and enables storage times of the ions in excess of
months. In principle this would allow to perform the complete measurement
with only one single pair of ions. Only to exclude systematic effects arising
from a possible residual contamination of the trap with other ions, we repeated
the measurement with two newly created pairs of ions.
\begin{figure}
\includegraphics[width=0.483\textwidth]{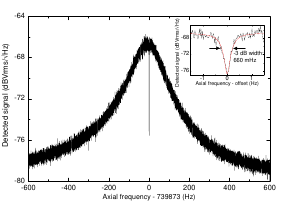}%
\caption{\label{Dip} Illustration of a typical dip spectrum for the
	determination of the proton axial frequency. The inset shows a zoom to the dip
	signal, together with our fitted lineshape model. For details see text.}
\end{figure}

Of the three independent Penning-trap eigenmotions, the tank circuit can only
detect the axial motion directly \cite{RevModPhys.58.233}. In order to
determine the cyclotron frequency of the stored ions, we measure the image
current the ions induce on the trap electrodes when oscillating with the axial
eigenfrequency $\nu_z$ of about $525~\text{kHz}$ for $^{12}\text{C}^{6+}$ and
$740~\text{kHz}$ for the proton, respectively. This tiny current is transformed
into a measurable voltage by a highly-sensitive superconducting tank circuit,
which is picked up by a low-noise cryogenic amplifier \cite{10.1063/1.4967493}.
Especially the proton with its low charge requires a high quality factor
($Q$-value) of the detection circuit to produce a sufficient signal. By
operating the trap and its electronics at cryogenic temperature ($4~\text{K}$),
the temperature of the tank circuit and with it also the kinetic energy of the
ions is reduced, which strongly suppresses systematic effects and increases
precision. Still, the low mass $m_{p}$ of the proton translates into relatively
large motional amplitudes for a given temperature, asking for exquisitely
well-defined electromagnetic trapping fields. Additionally, the finite kinetic
energy of the ion during the measurement causes a relativistic mass increase,
which is specifically strong for the light proton. Finally, even in a shimmed
superconducting magnet, the field is not perfectly homogeneous. In order to
guarantee that both ions are measured at exactly the same location, it is thus
important to use the identical trap voltage configuration for both ions. The
large charge-to-mass ratio mismatch of the two ions, and the respective large
voltages, would lead to shifts of the equilibrium position. These shifts are 
unavoidable and hard to control due to patch potentials. Combined with the
residual inhomogeneity of the magnetic field this would cause a systematic
error in the measured cyclotron frequency ratio. However, since it is currently
technically impossible to tune a high-$Q$ tank circuit over the required
frequency range, we have instead implemented for the first time two independent
tank circuits, fine-tuned to the exact ratio of the axial frequencies with a
voltage-variable capacitor. In this way we can keep the exact same voltage
setting for the measurement with respect to the two ions. 
The axial frequency can be determined from a fit to the noise dip that appears
as a unique signature of the ion when it is in thermal equilibrium with the
tank circuit (see Fig. \ref{Dip}). This dip has a $3~\text{dB}$-width of
$660~\text{mHz}$ for the proton and $1100~\text{mHz}$ for the carbon ion,
respectively. After an averaging time of three minutes, a fit allows to
determine the axial frequency with a precision of about $50~\text{mHz}$. 
The other two frequencies, the modified cyclotron frequency $\nu_+$ and the
magnetron frequency $\nu_-$, have to be determined by coupling them to the
axial motion with radio frequency drives on the motional sidebands
\cite{PhysRevA.41.312}. When driving the ion at the
\textquotedblleft red\textquotedblright{} axial-cyclotron
sideband at $\nu_+-\nu_z$, the axial
motion is dressed with the cyclotron state, leading to a splitting of the dip
signal into two dips (\textquotedblleft double-dip\textquotedblright{}),
from which the cyclotron frequency can be determined. For the
determination of the magnetron frequency a similar
technique is applied. From the three eigenfrequencies the free cyclotron
frequency can be calculated via the invariance relation
$\nu_c=\sqrt{\nu_+^2+\nu_z^2+\nu_-^2}$ \cite{PhysRevA.25.2423},
where $\nu_+ \approx 57~\text{MHz}$ for the proton and
$\nu_+ \approx 29~\text{MHz}$ for the carbon ion, respectively. Since the
modified cyclotron frequency dominates this relation, the relative precision
of its determination is of highest importance. For this reason, we utilize the
phase-sensitive PnA technique \cite{PhysRevLett.107.143003}, which allows
determining $\nu_+$ with highest precision and very low kinetic energy of the
ion, and thus low systematic frequency shifts. Furthermore, the influence of
temporal magnetic field fluctuations is reduced compared to the double-dip
technique, which requires longer measurement times. Both the PnA and the
double-dip technique are employed during the measurement campaign. While the
PnA technique is about an order of magnitude more precise for a single
measurement than the double-dip method, the comparison of the two techniques
allows for an important internal consistency check. Moreover, the PnA technique
is less prone to systematic shifts due to any imperfection in the lineshape.

\begin{figure}
\includegraphics[width=0.483\textwidth]{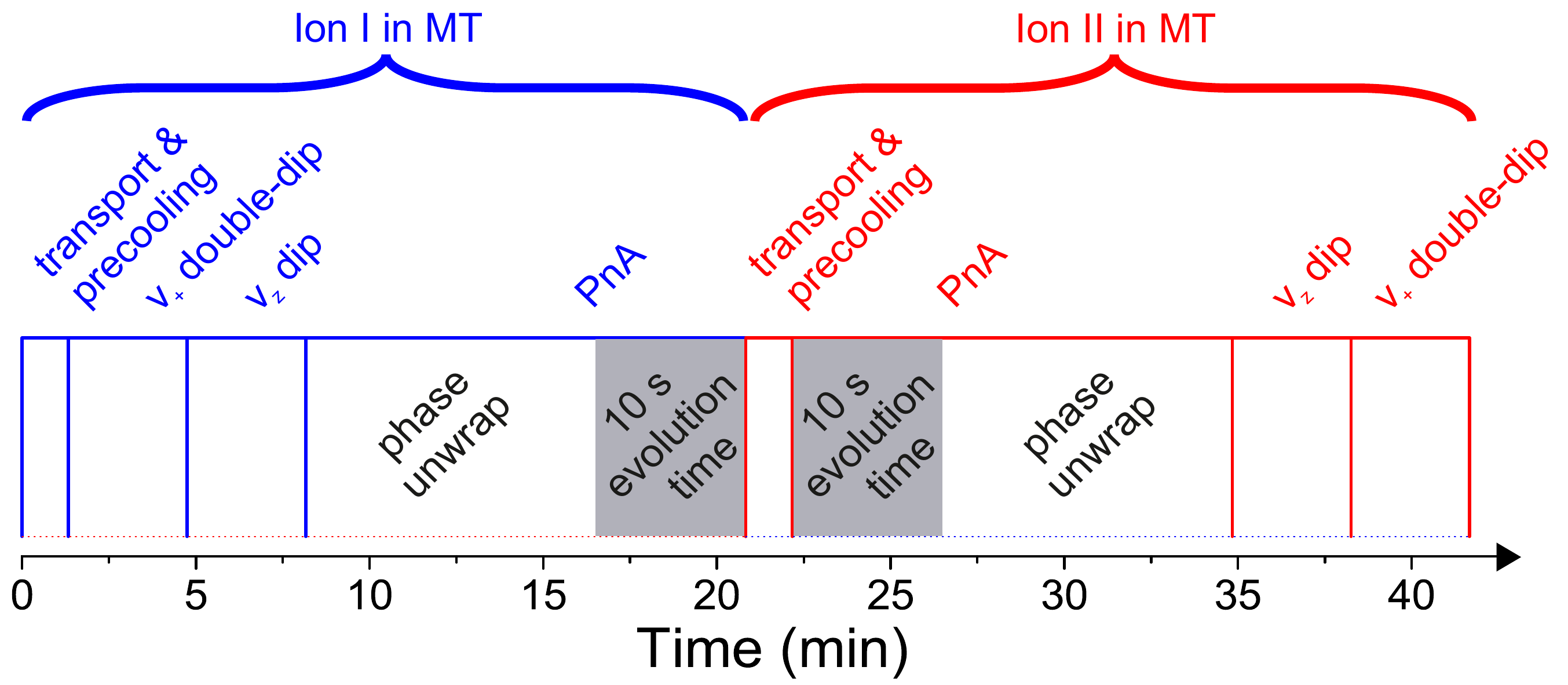}%
\caption{\label{zeitverlauf} Illustration of the measurement sequence. In the
beginning of each step, ion~I is chosen randomly to exclude linear magnetic
field drifts and systematic shifts arising from the measurement procedure.
Ion~I is then transported into the MT, and ion~II into the respective storage
trap. In any case, both storage traps are set to their nominal voltage to
prevent systematic influence on the equilibrium position of the measured ion
in the MT. After cooling the cyclotron motion as well as the axial motion,
$\nu_+$ and $\nu_z$ are measured with the dip methods, respectively. Ten
PnA cycles at different phase evolution times: six times $10~\text{ms}$,
$0.1~\text{s}$, $1~\text{s}$, $2~\text{s}$ and $5~\text{s}$ ensure the precise
determination of the initial phase and a proper phase unwrapping
\cite{PhysRevLett.107.143003}. Then four cycles of the PnA method are applied,
each with $10~\text{s}$ phase evolution time to determine $\nu_+$ with highest
precision. Finally ion~I is moved away and ion~II is loaded into the MT and its
frequencies are measured in reverse order. Each such cycle gives an individual
value for the mass ratio.}
\end{figure}

The sequence of a single measurement is illustrated in Fig.~\ref{zeitverlauf}.
At the beginning of each single ratio measurement, a random generator selects
the ion that gets measured first. This cancels the effect of a possible
systematic linear drift of the magnetic field. The first ion is then
transported to the MT and the other ion is shifted into its associated
storage trap. In both cases, the voltage in all three traps is set to
the same values, such that the electrostatic potential is truly identical.
Then the double-dip measurement of $\nu_+$ is performed, followed by an axial
frequency measurement. Subsequently, the modified cyclotron frequency is
measured again, this time with the PnA technique. Directly following the last
PnA cycle, the ions are swapped and the second half of the measurement cycle
starts with a PnA measurement on ion~II, such that the time in between the
cyclotron frequency measurements of the two ions is minimized. Finally, also
for the second ion the axial and cyclotron frequencies are measured using the
double-dip and single dip method, respectively. During one $43$-minute
measurement cycle the cyclotron frequency ratio is determined with a relative uncertainty
of $1.8\times10^{-10}$. 

The major systematic shift arises from the finite kinetic energy of the ions.
The thermal distribution of axial mode energies, which are thermalized with the
tank circuit, leads to a motion within the residual inhomogeneity of the
magnetic field. Owing to the axial symmetry of the motion, only even orders
of field moments are relevant, with the quadratic \textquotedblleft magnetic
bottle\textquotedblright{}
($\nicefrac{B_2}{B_0}=-7.2(4)\times 10^{-8}/\text{mm}^2)$ component being the
leading order contribution. The resulting shift of the modified cyclotron
frequency is particularly large for the proton, due to its low charge, and
a factor of six smaller for the $^{12}\text{C}^{6+}$ ion. To reduce the size
of this shift, the axial temperature of the ions is reduced below the ambient
cryogenic temperature to about $T=1.7(1.0)~\text{K}$ by means of electronic
feedback cooling \cite{UrsoFBCool}, which limits this shift to
$\nicefrac{\delta\nu_c}{\nu_c}=-44(28)~\text{ppt}$ for the proton. In order to 
determine this temperature we first perform sideband coupling of the cyclotron and 
the axial modes. Then a burst excitation of the cyclotron motion maps the initial axial 
amplitude distribution to the axial frequency via the residual magnetic bottle $B_2.$

\begin{table*} 
\begin{center}
\caption{Systematic shifts and their uncertainties for the individual cyclotron
frequencies and their ratio $R_0$. For details see text. The second column
gives the specific relative systematic shifts for the proton (left) and carbon
(right)
cyclotron frequencies, respectively. These values apply for the smallest
modified cyclotron radius of the proton ($9~\mu\text{m}$) and the carbon
ion ($14~\mu\text{m}$) during the phase evolution time of PnA. The values
in column three and four denote the relative systematic
shifts and their uncertainties for $R_0$
after the extrapolation to zero cyclotron energy for both ions. The lineshape
model shift only occurs for the first two ion pairs due to a slight offset
between the ion's and the detector's resonance frequencies and is therefore
not included in the total. The magnetron
frequency is not measured in each cycle, but due to its small value also
the resulting uncertainties of $\nu_c$ and $R_0$ are negligible within
the given limits.}\label{Sysshifts}
\begin{ruledtabular}
\begin{tabular}{l c c c}
 Effect & Rel. syst. shift of $\nu_c \left(10^{-11}\right)$ & Rel. syst. shift
 of $R_0 \left(10^{-11}\right)$ &  Uncertainty $\left(10^{-11}\right)$ \\
 $r_+^{\text{exc}}$ for p / $^{12}\text{C}^{6+}$ (\textmu$\text{m}$) & 9/14 & 0/ 0 & 0/0 \\ 
 \hline 
Image charge & 0.83/9.94 & 9.10 & 0.46\\
Image current & -0.14/-0.33 & -0.19 & 0.03\\
Residual magnetostatic inhomogeneity & 4.43/0.14 &-3.95  & 2.75  \\
Residual electrostatic anharmonicity & $\ll 0.01$/$\ll 0.01$ &
$\ll 0.01$ & $ \ll 0.01$  \\
Special relativity & 7.23/3.45  & -1.14 & 0.71 \\ 
Lineshape model\footnote{The typical value varies slightly between measurement sets due to different detunings of the axial resonators.} & -0.03/0.14 & 0.27 & 0.30\\
Magnetron frequency uncertainty & 0.01/0.06 & 0 & 0.06 \\
\hline
Total & 12.33/13.40 & 3.82 & 2.89  \\
\end{tabular}
\end{ruledtabular}
\end{center}
\end{table*}

\begin{figure}
\includegraphics[width=0.483\textwidth]{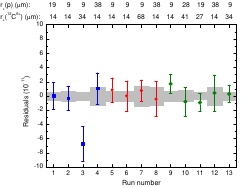}
\caption{\label{Results2} Residuals of the 3-parameter ($R_0,a,b)$ fit with
$R_i=R_0+a \left(A_{t,i} (p)\right)^2 + b \left(A_{t,i}
(^{12}\text{C}^{6+})\right)^2$ to the measured frequency ratios $R_i$. Here,
$A_{t,i}(p)$ and $A_{t,i}(^{12}\text{C}^{6+})$ denote the effective dipole
excitation strength, the product of excitation amplitude and time for the
modified cyclotron motion of the proton and the carbon ion, respectively.
The shown modified cyclotron radii can be calculated via $r_{+,i}=\kappa
\times A_{t,i}$, where the parameter $\kappa$ is extracted from $a$ and
$b$, respectively. This is checked via an independent calibration of the
amplitudes by means of a frequency shift due to the residual magnetic
inhomogeneity and shows good agreement. The grey area indicates the prediction
interval of the fit, the error bars indicate the statistical uncertainty of the
individual measurement. The agreement for the complete range of modified
cyclotron energies indicates the validity of our model of systematics. The
data set consists of three separate ion pairs, indicated by the color of the
data points. The agreement of the data sets renders an influence of
parasitically trapped ions or electrons improbable.}
\end{figure}

Apart from the axial temperature also the cyclotron energy after the excitation
within the PnA cycle contributes to the systematic shifts. Here, the mass
increase due to special relativity is dominant and mainly affecting the proton,
resulting in a shift of $\nicefrac{\delta\nu_c}{\nu_c}=-72(8)~\text{ppt}$ for the
proton. By an extrapolation using varying excitation amplitudes it can be
corrected for (see Fig.~\ref{Results2}). The absence of any non-statistical
jitter in the corrected ratios gives us great confidence in the validity of our
systematics model.

\begin{figure}
\includegraphics[width=0.483\textwidth]{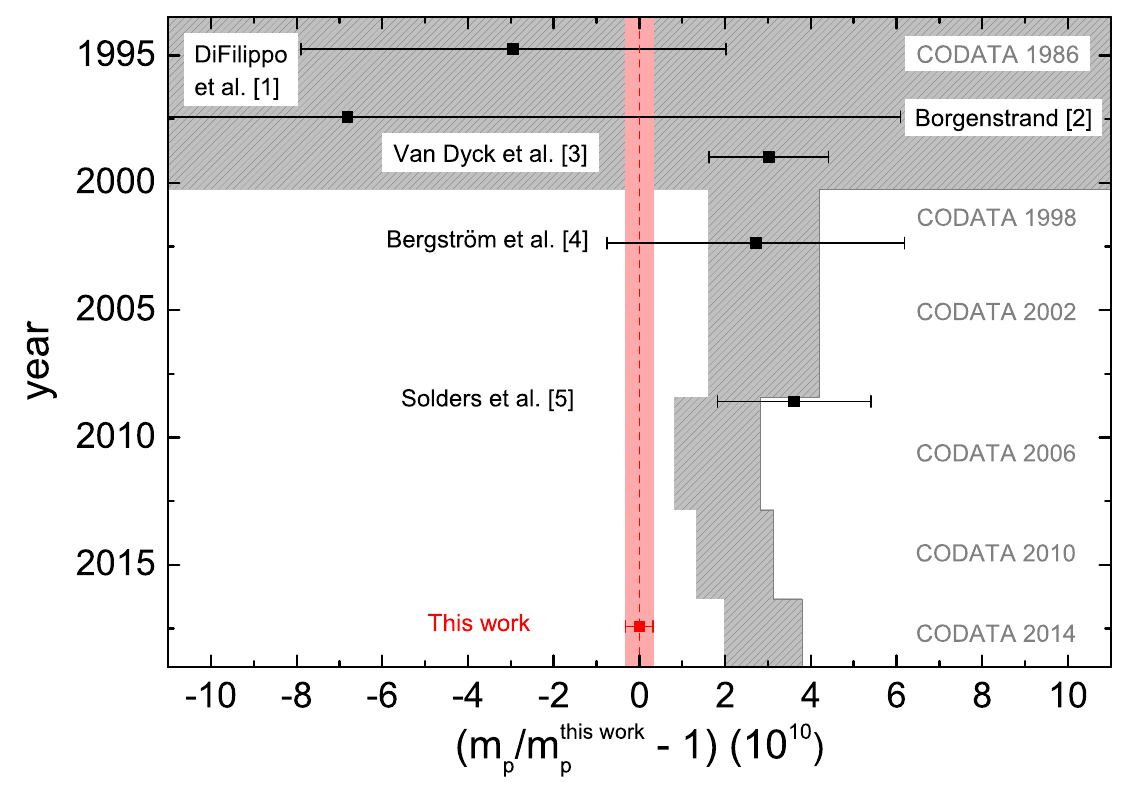}%
\caption{\label{History} Comparison of our result to previous values for the
proton's atomic mass. Mainly two Penning-trap experiment's contribute to the
literature value, the UW-PTMS at the University of Washington
\cite{doi:10.1063/1.57450} and the SMILETRAP spectrometer in Stockholm
\cite{Borgenstrand436573,1402-4896-66-3-002,Solders232785}. Our value disagrees
with the latest CODATA value at a level of $3.3$ standard deviations.}
\end{figure}

Additionally to the above-mentioned energy dependent
effects, the major systematic shift arises from the interaction of the ion with
the trap electrodes. Due to the axial symmetry of the trap, the ion's image
charges in the electrodes mainly produce an outward force on the ion in the radial
direction. This shifts both the modified cyclotron frequency and the magnetron
frequency, but leaves the axial frequency unaffected apart from a tiny
contribution due to the slits between the trap electrodes. For that reason,
the shift does not cancel out when using the invariance relation of the three
motional frequencies and has to be calculated and corrected. However, compared
to previous measurements \citep{Nature2014}, our new and larger trap helps to
diminish this effect, which scales as 
\begin{equation}
\frac{\delta\nu_c}{\nu_c}=- C_{\text{IC}}\frac{m}{8 \pi
\epsilon_0 r_0^3 B_0^2},
\end{equation} 
where $r_0=5~\text{mm}$ denotes the trap radius and $B_0=3.76~\text{T}$. The
coefficient $C_{\text{IC}}$ depends on the exact dimensions of the electrodes
and their gaps and has been determined numerically to $C_{\text{IC}}=1.97(10)$
\cite{Marc}. Due to the dependence on the ion's mass, this shift affects mostly
the carbon ion, where it amounts to $\nicefrac{\delta\nu_c}{\nu_c}=
-99~\text{ppt}$. The uncertainty of this shift is given by the manufacturing
precision of the trap electrodes of $\pm 10~\mu\text{m}$, allowing to correct
the shift to better than $5~\text{ppt}$.
Apart from the immediate interaction of the ion with its image charges,
depending on the impedances of the electronics attached to the electrodes,
also the resulting image current interacts with the ion. Close to their
respective resonances, high-$Q$ tank circuits boost this interaction and
can lead to sizeable frequency pulling \cite{PhDNatarjan}. To control and
reduce this effect, our cyclotron tank circuits, which are used for faster
identification of the ions, are detuned by several line-widths during the
measurement. The residual shift is estimated to
$\nicefrac{\delta\nu_c}{\nu_c}=3.3(2)~\text{ppt}$ for the carbon ion and even
smaller for the proton. The same effect in the axial motion is taken care
of by our lineshape model, which corrects for the frequency
pulling. However, due to imperfect knowledge of the resonator parameters
and a slight off-resonance of the ions in some of our measurements this
contributes to systematic uncertainties. Our new seven-electrode cylindrical
trap enables the adjustment of the electrostatic potential in a way that
contributions to the systematic error budget are completely negligible.
This allows us to parametrically amplify the ion motion at the end of the
PnA cycle to large enough amplitudes for achieving a sufficient signal to
apply PnA to the proton, which is challenging due to its low charge.
A summary of all shifts is listed in Table~\ref{Sysshifts}.
From our measurements and the extrapolation to zero cyclotron energy we
obtain (see Fig.~\ref{Results2}):
\begin{equation}
R_0~\equiv~\frac{\nu_c(^{12}\text{C}^{6+})}{\nu_c(p)}\vert_{\text{stat}}=
0.503\,776\,367\,643\,1(77).
\end{equation}
Applying all corrections according to Table~\ref{Sysshifts} we arrive at:
\begin{equation}
R_{\text{final}}\vert_{\text{stat,sys}}~=~0.503\,776\,367\,662\,4 (77)(146).
\end{equation}
By correcting for the mass of the missing electrons and their respective
binding energies in the $^{12}\text{C}^{6+}$ ion, taken from \cite{NIST_IP},
we can relate the mass of the carbon ion to the atomic mass unit as shown in
Eq. (\ref{iE}). Finally, using
$m_{p}=R_{\text{final}}\,m(^{12}\text{C}^{6+})/6$ we calculate
the proton mass in atomic mass units:
\begin{equation}
m_{p}=1.007\,276\,466\,583(15)(29)~\text{u}.
\end{equation} 
Here, the two numbers given in brackets are the statistical and systematic
uncertainties of the measurement, respectively. Thus, our value of $m_{p}$
has a relative precision of $32~\text{ppt}$, which is three times more precise
than the current CODATA value (Fig.~\ref{History}) \cite{RevModPhys.88.035009}
but shows a deviation from the literature value by more than three standard
deviations.

As an additional cross-check we can use the double-dip measurements of
$\nu_+$ instead of the PnA, which yields:
\begin{equation}
R_{\text{DD}}\vert_{\text{stat,sys}}=0.503\,776\,367\,66 (3)(5). 
\end{equation}
The result from the double-dip data is in excellent agreement with the PnA
result, however by a factor of around four less precise.
To further confirm the measured $3.3~\sigma$ deviation we conducted a sequence
of cross-check mass measurements.
To this end, we have performed similar measurements as above, however replacing
the highly charged carbon ion by $^{16}\text{O}^{8+}$.
Using the respective version of Eq.~(\ref{iE}) and associated ionization
energies~\citep{NIST_IP} for $^{16}\text{O}$ and our new value for $m_p$ we
obtain
\begin{equation}
m\left(^{16}\text{O}\right)=15.994\,914\,619\,24 (54)(43)(53)~\text{u,} 
\end{equation}
where the last bracket contains the uncertainty arising from our measured
proton mass. The relative uncertainty of this result is $5.4\times 10^{-11}$,
and with an $0.4~\sigma$ deviation it is in excellent agreement with the
literature value of the AME2016
$m\left(^{16}\text{O}\right)= 15.994\,914\,619\,60 (17)~\text{u}$ \cite{1674-1137-41-3-030002}.
Finally, we performed a comparison of the cyclotron frequency ratio of
$^{12}\text{C}^{3+}$ and $^{12}\text{C}^{6+}$.
The measured mass agrees to the calculated one within $0.5~\sigma$ with
an uncertainty of $1.1\times 10^{-10}$, where the relative systematic
uncertainty is only $7\times 10^{-12}$.

Using our result, the proton-electron-mass ratio can be determined with
a relative precision of $43~\text{ppt}$, where the uncertainty arises nearly
equally from the proton and the electron mass. This is a factor of two more
precise compared to the current value \cite{RevModPhys.88.035009}:
\begin{equation}
\nicefrac{m_{p}}{m_{\text{e}}}=1\,836.152\,673\,346(81).
\end{equation}
The shifted proton mass also impacts the $^3\text{He}$ \textquotedblleft mass
puzzle\textquotedblright{} \cite{PhysRevLett.114.013003,0026-1394-52-2-280},
which indicated a possible inconsistency of the existing determination of the
mass of the $\text{HD}^+$ molecule compared to $^3\text{He}^+$. The
inconsistency of $4~\sigma$ is reduced by a factor of around two using
our measurement result. By applying our measurement scheme also with the
deuteron, we will be able to further address the $^3\text{He}^+$ inconsistency.
Furthermore, $m_{p}$ affects the atomic mass of the
neutron~\citep{0026-1394-52-2-280}, but results in a shift of smaller than
$1~\sigma$, due to the dominant uncertainty in the deuteron's binding energy.
The influence on the Rydberg constant $R_{\infty}$ \cite{Karshenboim2016} is
currently small, since its error is dominated by the charge radius
of the proton. However, the more precise value for $R_{\infty}$ that could be
extracted from the muonic hydrogen experiment once the proton radius puzzle
\citep{Pohl2010} can be resolved will be significantly influenced by our result.

In summary we performed the most precise measurement of the atomic mass of the
proton. Our measurement is a factor of three more precise compared to the
current literature value, however shifted by about three standard deviations.
In a set of carefully conducted cross-check measurements we have confirmed a
series of other literature values and were not able to track any yet uncovered
systematic effects imposed by our method. Combined with the independently
measured electron mass this measurement yields a factor of two more precise
proton-electron-mass ratio, too.

The main systematic limitation of our measurement is given by the residual
quadratic magnetic field component combined with the finite axial motion
amplitude of the ions. In the next phase of our experiment we plan to
significantly improve on this limitation by compensating the first and
second order magnetic inhomogeneities with a dedicated set of in-situ
superconducting magnetic shims. Additionally, common-mode magnetic field
fluctuations will be canceled by simultaneous phase-sensitive measurements
in the RT and MT, allowing for significantly longer measurement times and
correspondingly a lower statistical uncertainty.
\section{Acknowledgements}
We want to thank Marc Schuh for the calculation of the image charge shift
in our trap. This work was supported by the Max Planck Society, the EU
(ERC grant no 290870; MEFUCO), the International Max Planck Research School
for Quantum Dynamics in Physics, Chemistry and Biology (IMPRS-QD) and the
RIKEN FPR Funding as well as the RIKEN Incentive Research Project Program.
\bibliography{ProtonMass2017sources_17052017}

\end{document}